\documentclass[prl,twocolumn,nofootinbib, preprintnumbers, superscriptaddress]{revtex4-1}

\usepackage{amsmath,amssymb,amscd,simplewick}
\usepackage{listings}
\usepackage{dsfont}
\usepackage{slashed}
\usepackage{color}
\usepackage{ulem}

\usepackage{graphicx}
\usepackage{epstopdf}
\usepackage{subfigure}
\usepackage{epsfig}

\usepackage{xcolor}
\usepackage[colorlinks=true,
            linkcolor=blue,
            urlcolor=blue,
            citecolor=green,          
            bookmarks=true,
            bookmarksnumbered=true,
            breaklinks=true,
            pdfpagemode=Fullscreen,
            pdfstartview=FitBH]{hyperref}

\hypersetup{pdfauthor = {Shao-Feng Ge},
	     pdftitle = {}, 
	     pdfsubject = {}, 
             pdfkeywords = {}, 
	     pdfcreator = {LaTeX with hyperref package},
	     pdfproducer = {dvips + ps2pdf} }
 
\usepackage{float}
\definecolor{gesfpurple}{rgb}{0.47,0.19,0.42}

\definecolor{gesflanse}{rgb}{0.00,0.50,0.50}

\definecolor{gesfblue}{rgb}{0.08,0.42,0.76}

\definecolor{gesfred}{rgb}{1,0,0}

\definecolor{gesfwhite}{rgb}{1,1,1}

\definecolor{gesfblack}{rgb}{0,0,0}

\newcommand{\geqn}[1]{\hypersetup{linkcolor=blue}Eq.~(\ref{#1})\hypersetup{linkcolor=blue}}
\newcommand{\gfig}[1]{{\hypersetup{linkcolor=violet}Fig.~\ref{#1}\hypersetup{linkcolor=blue}}}
\newcommand{\gtab}[1]{{\hypersetup{linkcolor=gesflanse}Tab.~\ref{#1}\hypersetup{linkcolor=blue}}}

\newcommand{\fpar}{\slashed \partial}

\newcommand{\cnb}{C$\nu$B}

\usepackage{multirow}

\graphicspath{{figs/}}

\begin{document}

\title{Parity Violation and Chiral Oscillation of Cosmological Relic Neutrinos}
\author{Shao-Feng Ge}
\email{gesf@sjtu.edu.cn}
\affiliation{Tsung-Dao Lee Institute, Shanghai Jiao Tong University, Shanghai 200240, China}
\affiliation{School of Physics and Astronomy, Shanghai Jiao Tong University, Shanghai Key Laboratory for Particle Physics and Cosmology, Shanghai 200240, China}
\author{Pedro Pasquini}
\email{ppasquini@sjtu.edu.cn}
\affiliation{Tsung-Dao Lee Institute, Shanghai Jiao Tong University, Shanghai 200240, China}

\begin{abstract}
The conventional derivation of neutrino oscillation
treats neutrino mass eigenstate as
plane wave with an overall evolution phase. Nevertheless,
due to the intrinsic parity-violating nature of weak
interactions, only the left-chiral
neutrino can be produced as initial condition.
On the other hand, the neutrino mass term connects the
left-chiral component to the right-chiral one and
unavoidably leads to generation of the
later through oscillation. This chiral oscillation
has significant consequences on the detection of the
cosmological relic neutrinos. The event rate is
reduced by a factor of 2 than the conventional prediction.
\end{abstract}
\maketitle

{\it Introduction} --
The neutrino oscillation \cite{Pontecorvo:1957cp,MNS,P67}
was established by the atmospheric \cite{Fukuda:1998mi}
and solar \cite{solarNu,Ahmad:2001an} neutrino experiments.
It indicates that the active neutrinos are massive,
rather than massless particles as assumed in the
Standard Model (SM) of particle physics \cite{SM}.
This is the first new physics beyond the SM.

Although neutrino is a fermion that is described by
spinor \cite{Giunti:2007ry}, the conventional
derivation \cite{derivation}
 simply takes each neutrino
mass eigenstate $\nu_i$ as a plane wave,
$\nu_i(x) = e^{- i p \cdot x} \nu_i$, where
$p$ and $x$ are the 4-momentum and coordinates,
to describe the
neutrino oscillation, completely neglecting the
chiral nature of neutrinos. In this formalism,
the neutrino oscillation happens
among different flavors.
Nevertheless, this approach is far from being enough.

The oscillation effect is intimately related to the
tiny neutrino masses. To make the effect of the
tiny mass differences explicit, the neutrino
propagation and oscillation should happen over large
enough distance. Although we infer nonzero neutrino
masses from neutrino oscillations, it is difficult
to make the chiral properties of neutrinos manifest
due to the smallness of neutrino masses. Neutrinos
can be effectively described by the two-component
theory \cite{Lee:1957qr}. Practically, the tiny neutrino masses
also make it very difficult to distinguish Dirac
and Majorana neutrinos \cite{Kayser:1982br}.

The incorporation of the spinor nature of neutrinos
in the oscillation leads naturally to what is called
the chiral oscillation
\cite{DeLeo:1996gt,Bernardini:2005wh,Bernardini:2007ew,Bernardini:2010zba}
and also gives rise to the interesting phenomena of
neutrino-antineutrino oscillation for Majorana fields
\cite{Pontecorvo:1957cp,Schechter:1980gk,Li:1981um,deGouvea:2002gf,Xing:2013ty,Langacker:1998pv}.
On top of that, another type of phenomena can appear
in the relativistic theory of fermion propagators,
such as helicity flip and the coupling with magnetic
fields \cite{Popov:2019nkr,Pustoshny:2020fqv}.

To make the spinor nature of neutrino oscillation
explicit, we need low energy neutrino flux.
The cosmological relic neutrinos from the early
Universe is a perfect test ground for this purpose.
Their energy is typically of the order of meV as
predicted by the Cosmological Standard Model
\cite{Dolgov:2002wy}. In this paper, we evaluate
the phenomenological consequences of spinor oscillation
on the cosmic neutrino background (C$\nu$B) for both
Dirac and Majorana types.

{\it Chiral Oscillation}
--
Two things need to be incorporated into the description
of neutrino oscillation. First, the neutrino interactions
in the SM violate parity \cite{Lee:1956qn}. In both the
neutral and charged current interactions, only
left-chiral neutrino appears while the right-chiral one is
completely absent. It is then not reasonable to take the
neutrino mass eigenstate as a whole particle by treating
its evolution with just an overall phase $e^{- i p \cdot x}$ 
without considering its spinor nature.
Instead, only the
left-chiral neutrino can be created at the production point.

During the propagation, the left-chiral neutrino oscillates
to its right-chiral counterpart due to the connection
established by the mass term. A derivation with operator
projections can be found in
\cite{Bernardini:2005wh,Bernardini:2007ew,Bernardini:2010zba}. 
Here we provide a more
intuitive derivation. For a neutrino with fixed momentum
\begin{equation}
  \nu(t, {\bf x})
=
  \nu(t) e^{i {\bf p} \cdot {\bf x}}
=
\left\lgroup\begin{matrix}
  \varphi(t) \\
  \chi(t)
\end{matrix}
\right\rgroup
 e^{i {\bf p} \cdot {\bf x}} \,,
\label{eq:ansatz}
\end{equation}
we parametrize the time dependence as $\varphi(t)$ and
$\chi(t)$. The Dirac equation written explicitly in terms
of chiral components looks like,
\begin{equation}
\left\lgroup
\begin{matrix}
  m & - i \partial_t  + \boldsymbol\sigma \cdot {\bf p} \\
  -i \partial_t - \boldsymbol \sigma \cdot {\bf p} & m
\end{matrix}
\right\rgroup
\left\lgroup
\begin{matrix}
  \varphi(t) \\
  \chi(t)
\end{matrix}
\right\rgroup
=
  0 \,.
\label{eq:DiracEq}
\end{equation}
Without mass, the left- and right-chiral
components disentangle from each other. In other
words, the mass makes it possible to express
$\chi(t)$ in terms of $\varphi(t)$,
\begin{equation}
  \chi(t) 
= 
  \frac 1 m
\left[
  i \partial_t
+ {\bf p} \cdot \boldsymbol \sigma
\right] \varphi(t).
\label{eq:chi(t)}
\end{equation}
Only one of $\chi(t)$ and $\varphi(t)$ is independent.
The solution can be generally parametrized as
\begin{equation}
  \varphi(t)
=
  C_1 \cos (Et) + C_2 i \sin (Et),
\end{equation}
where $E$ is the neutrino energy while $C_1$
and $C_2$ denote two spinors. Substituting
$\varphi(t)$ into \geqn{eq:chi(t)}, the
corresponding $\chi(t)$ can be obtained,
\begin{eqnarray}
  \chi(t)
& = &
  \frac 1 m
\left[
  \left(
   {\bf p} \cdot \boldsymbol \sigma C_1
  - C_2 E
  \right) \cos (E t)
\right.
\nonumber
\\
&&
\left.
\hspace{-2mm}
+ i
\left(
- E C_1
+ {\bf p} \cdot \boldsymbol \sigma C_2
\right) \sin (E t)
\right] \,.
\end{eqnarray}

The initial condition requires $\chi(t = 0) = 0$, establishing
correlation among
the two spinors, $C_2 = ({\bf p} \cdot \boldsymbol
\sigma / E) C_1$. For the production of a left-chiral neutrino,
the spinor $C_1$ is exactly the initial $\varphi(0)$.
A similar calculation can be performed for an initial right-chiral antineutrino.
The oscillating neutrino then becomes
\begin{equation}
  \nu(t)
=
\left[
  \cos (E t)
- \frac i E
\left(
  2 c {\bf p} \cdot {\bf S}
+ m \gamma_0
\right)
  \sin (Et)
\right] \nu_c,
\label{eq:spinor_LH_time}
\end{equation}
where $c = \mp$ for left- and right-chiral
components $\nu_c = \nu_{L,R}$, respectively,
for generality. As we can see, switching the
chirality in \geqn{eq:DiracEq} changes the sign
in front of the momentum.

The spin operator is defined as,
${\bf S} \equiv \frac 1 2 \mbox{diag}\{\boldsymbol\sigma, \boldsymbol \sigma\}$.
Consequently,
$2 {\bf p} \cdot {\bf S} \nu^h = |{\bf p}| h \nu^h$
extracts the helicity eigenvalue $h \equiv \pm$.
Note that $\gamma_0$
does not change spinor helicity. Although the neutrino
spinor oscillates among different chiralities,
its helicity is conserved. A more explicit form is
\begin{equation}
  \nu(t)
=
\left\{
\left[
  c_E
- i c h \frac {|{\bf p}|} E s_E
\right]
- \frac m E s_E \gamma_0
\right\} \nu_c,
\label{eq:psi(t)-chiral}
\end{equation}
where we have defined
$[c_E, s_E] \equiv [\cos(E t), \sin(E t)]$
for simplicity. The first term of \geqn{eq:psi(t)-chiral}
conserves chirality while the second changes it.
In the massless limit, $m \ll E$, the evolution
reduces to just an chirality-conserving overall
phase, $\nu^h_c(t) = e^{- i c h E t} \nu^h_c$. On the other
hand, the non-relativistic limit gives a maximal
oscillation between the left- and right-chiral
components, $\nu(t) = c_E \nu_L - s_E \nu_R$,
where $\nu_R \equiv \gamma_0 \nu_L$. Note that this
oscillation happens even with only one flavor of neutrino,
\begin{subequations}
\begin{align}
  P_{\nu_L\rightarrow \nu_L}
& =
  |c_L|^2
=
  \cos^2(Et)
+  \frac {|{\bf p}|^2} {E^2} \sin^2(Et),
\\
  P_{\nu_L\rightarrow \nu_R}
& =
  |c_R|^2
=
  \frac {m^2}{E^2} \sin^2(Et).
\label{eq:LR_probs}
\end{align}
\end{subequations}

In the SM, this Majorana spinor can be composed as,
$\nu_M = \nu_L + \mathcal C \overline{\nu_L}^T$.
The Dirac equation of Majorana fields,
$i \fpar \nu_L = m \mathcal C \overline{\nu_L}^T$
\cite{Giunti:2007ry}
and equivalently
$i \fpar \mathcal C \overline{\nu_L}^T = m \nu_L$
can group into the same form as \geqn{eq:DiracEq}.
And the initial left-chiral Majorana neutrino in the
relativistic limit would unavoidably have both
chiralities in the non-relativistic limit
\cite{Long:2014zva}. The chiral oscillation
\geqn{eq:spinor_LH_time}
applies for both Dirac and Majorana neutrinos.
With these taken into
consideration, the Majorana condition,
$\nu^c_M(t) = \nu_M(t)$, holds for chiral
oscillation at an arbitrary time.
Valuable discussions can be found in \cite{Dvornikov:2010dc,Esposito:1997fy,
Esposito:1998hh,Fukugita:2003en}.

{\it Chiral Evolution of Cosmic Neutrinos} --
Since the chiral oscillation is induced by the
nonzero mass, its effect is most significant at
the nonrelatistic limit. However, the
neutrino masses are very tiny, being at the sub-eV
scale. It is very difficult to find low-energy
neutrino fluxes to test
the chiral oscillation with parity violation.

Fortunately, the cosmic relic neutrinos created
at the early stage of the expanding Universe have
cooled down to low enough energy and hence can
serve as a perfect place for testing the chiral
oscillation.

The evolution of the neutrino states can be described
by the Boltzmann equation \cite{deSalas:2016ztq},
\begin{equation}
  \mathbb L(\rho)
\equiv
\left(
  \partial_t
- H {\bf p} \cdot \nabla_{\bf p}
\right) \rho
=
- i \left[ \mathbb H, \rho \right] \,,
\label{eq:LHrho}
\end{equation}
where $H$ is the Hubble constant and $\mathbb H$
is the Hamiltonian,
\begin{equation}
  \mathbb H
=
  {\bf p} \cdot (\gamma_0 \boldsymbol \gamma)
+ m \gamma_0
=
\left\lgroup
\begin{matrix}
  h |{\bf p}| & m \\
  m & - h |{\bf p}|
\end{matrix}
\right\rgroup \,,
\end{equation}
in the chiral basis. The left-land side of the
Boltzmann equation is the Liouville operator
$\mathbb L$ which incorporates the Hubble expansion.
On the other hand, the right-hand side is not
from the collision that a Boltzmann equation
usually describes but from the chiral oscillation.
With chiral decomposition, the density matrix also
becomes a $2 \times 2$ matrix,
\begin{equation}
  \rho
\equiv
\left\lgroup
\begin{matrix}
  \rho_{LL} & \rho_{LR} \\
  \rho^*_{LR} & \rho_{RR}
\end{matrix}
\right\rgroup \,,
\label{eq:density_matrix}
\end{equation}
where $\rho_{LL}$ and $\rho_{RR}$ stand for the
fraction of the left- and right-chiral
components, respectively.

To make it more explicit, we may regroup the density
matrix elements,
$\rho_\pm \equiv \rho_{LL} \pm \rho_{RR}$,
$\rho_R \equiv \mathbb R(\rho_{LR})$ and
$\rho_I \equiv \mathbb I(\rho_{LR})$.
The total probability $\rho_+$ is conserved,
$\mathbb L[\rho_+] = 0$, since
its contribution to the $\rho$ matrix commutes
with the Hamiltonian. The remaining $\rho_-$,
$\rho_R$, and $\rho_I$ are related to each other.

Since the Hubble constant is typically smaller
than the neutrino mass, the solution can be
obtained by series expansion of
$\epsilon \equiv H / m$. The leading order result
is,
\begin{subequations}
\begin{eqnarray}
  \rho_-
& = &
  1
- \frac{2m^2}{E^2}\sin^2(Et),
\\
  \rho_R
+ i \rho_I
& = &
  \frac {h m |{\bf p}|}
        {E^2}
  \sin^2(Et)
+ i \frac m E \sin(2 E t).
\end{eqnarray}
\label{eq:rho-solutions}
\end{subequations}
The next order expansion solution is
$\lesssim 1\%$ in the limit $|{\bf p}_{0}|\ll m$
at the current time.

For the typical $m = \mathcal O(10^{-3} \sim 10^{-2})$\,eV
neutrinos, the chiral oscillation is fast enough,
in comparison to the big time difference between
neutrino decoupling to now,
$t_{0}- t_{\rm d} \sim 6 \times
10^{38} {\rm MeV}^{-1}$.
The averaged result becomes,
\begin{equation}
  \langle\rho_-\rangle 
=
  1 - \frac{m^2}{E^2}, 
\quad 
  \langle\rho_R\rangle 
= 
  \frac{hm|{\bf p}|}{E^2}, 
\quad
  \langle\rho_I\rangle 
=
  0\,.
\label{eq:rho_1nu}
\end{equation}
For neutrinos today with very small redshift,
its momentum is much smaller than its mass
($|{\bf p}_{0}|\ll m$) or equivalently $E \approx m$,
leading to $\rho_- = 0$. Hence, the relic neutrinos today
have equal contributions from the left- and right-chiral
components, $\rho_{LL} = \rho_{RR} = 1/2$,
even though at the beginning therewas only left-chiral
component. Since $\rho_I$ averages out, we only need
to consider $\langle \rho_- \rangle$ and
$\langle \rho_R \rangle$ in the following discussions.

\begin{figure*}[t]
\centering
\includegraphics[scale = 0.3]{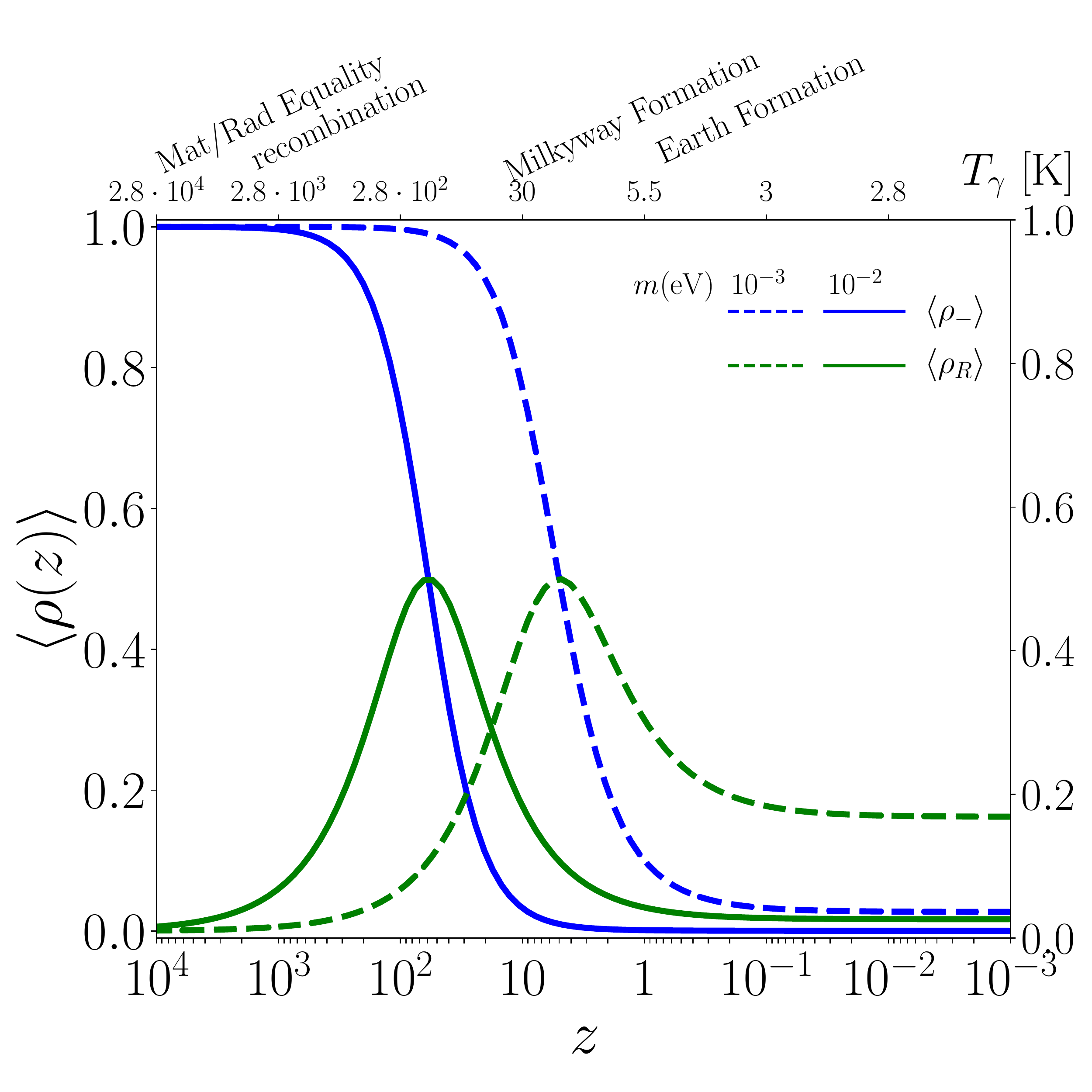}
\hspace{5mm}
\includegraphics[scale = 0.3]{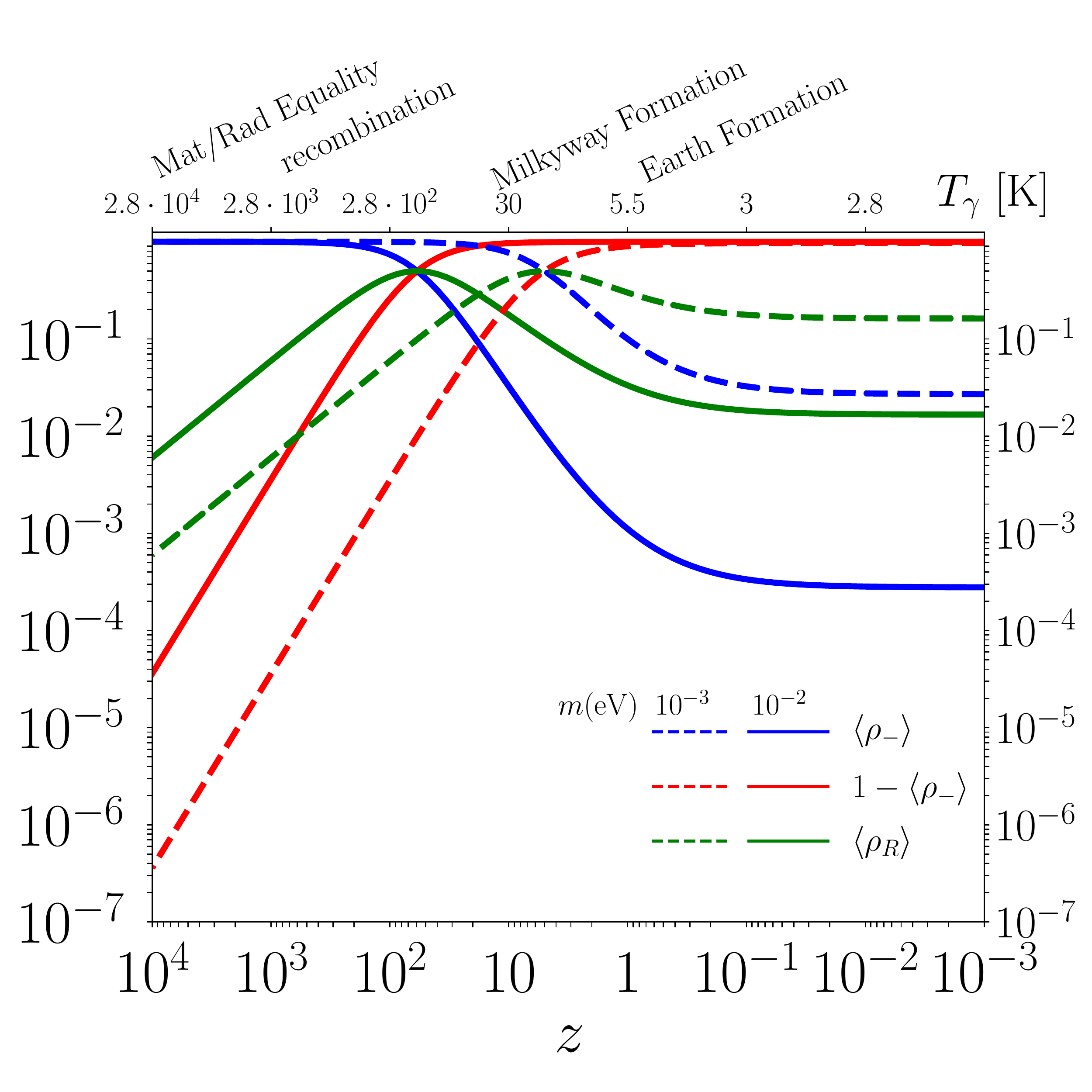}
\caption{Evolution of the chiral density matrix elements
$\langle\rho_-\rangle$ (blue) or $1-\langle\rho_-\rangle$
(red), and $\rho_R = \mathbb R[c_Lc_R^*]$ (green) as
functions of the redshift $z$ for $m = 10^{-2}$\,eV
(solid) and $m= 10^{-3}$\,eV (dashed) with
$|{\bf p}_d|= 1$\,MeV. The redshift at neutrino
decoupling is taken to be $z_d \sim 6 \times 10^9$.}
\label{fig:rho_vrs_redshift}
\end{figure*}
The results are shown in \gfig{fig:rho_vrs_redshift}
as a function of the redshift $z$ or CMB temperature
$T_\gamma$ for $m = 10^{-2}$ eV (solid line) and $m= 10^{-3}$
eV (dashed line),
together with a typical decoupling momentum
$|{\bf p}_d| = 1$\,MeV which
corresponds to $|{\bf p}_0| \sim 1.66 \times 10^{-4}$\,eV
at the present time.
The neutrino decoupling happened at redshift
$z_{\rm d} \sim 6\times10^9$. The current redshift
is $z_0=0$ and temperature around
$2.75$ K $ \sim 2.4\times 10^{-4}$\,eV.
Due to the Universe expansion, the particle momentum
gets redshifted as $|{\bf p}(z)| = (1 + z)|{\bf p}_0|$.
The solutions \geqn{eq:rho-solutions} are actually
functions of the redshift $z$.
The behavior of the curves of $\langle\rho_-\rangle$ and
$\langle\rho_R\rangle$ can be understood analytically,
\begin{equation}
  1 - \langle \rho_- \rangle
\approx
  \frac {\hat \rho^2_0}
        {(1+z)^2 + \hat \rho^2_0},
\quad
  \langle \rho_R \rangle
\approx
  \frac{\hat \rho_0 (1+z)}
       {(1+z)^2 + \hat \rho^2_0}\,,
\end{equation}
where $\hat \rho_0 \equiv m / |{\bf p}_0|
\sim 6\times 10^4 \left(\frac m {\rm eV}\right)$.
Notice that $\langle\rho_R\rangle$ has a maximum
at $z_{\rm peak} \equiv \hat\rho - 1$, where
$\langle \rho_R \rangle = \langle \rho_- \rangle = 1/2$
that also appears as a crossing point between the
$1 - \langle \rho_- \rangle$ and $\langle \rho_R \rangle$
curves. Since the redshift has to be positive,
this can only occur if $\hat \rho_0 \geq 1$ or
$m \geq|{\bf p}_0| \sim 1.66 \times 10^{-4}$\,eV.
For comparison, we show in \gfig{fig:rho_vrs_redshift}
a solid line for $m = 10^{-2}$\,eV and a dashed line
for $m = 10^{-3}$\,eV. Consequently, there is a
shift almost one order of magnitude in the position
of the peak between these two cases. The solid
line peaks at $z_{\rm peak} = 599$ while the dashed one
at $z_{\rm peak} = 59$.

We may further explore the behaviors at the large
redshift $z \gg \hat \rho_0$,
\begin{equation}
  1
- \langle \rho_- \rangle
\approx
  \frac {\hat \rho^2_0}{z^2}
=
  \frac{m^2}{|{\bf p}_0|^2 z^2},
\quad
  \langle \rho_R \rangle
\approx
  \frac {\hat \rho_0} z
=
  \frac m {|{\bf p}_0| z}\,.
\end{equation}
To make the scaling behaviors more explicit,
we show log-scale plots in the right plot of
\gfig{fig:rho_vrs_redshift}. The slope of the
red $1 - \langle \rho_- \rangle$ curves for large
$z$ is $-2$ while the one for the green
$\langle \rho_R \rangle$ curve is $-1$.
Since it is a log-log plot, the intersection of
the curves with the y-axis is determined by the
relative size of $m$ and $|{\bf p}_0|$. The $m^2$
dependence induces two orders of magnitude difference
between the two intersections,
$1 - \langle \rho_-(z=10^4) \rangle
\sim 3.6\times10^{-5}$ for $m = 10^{-2}$\,eV and
$3.6\times10^{-7}$ for $m = 10^{-3}$\,eV on
the y-axis. On the other hand, the green curve has
only one order of magnitude difference,
$\langle \rho_R (z=10^4) \rangle \sim 6 \times 10^{-3}$
for $m = 10^{-3}$\,eV and $6 \times 10^{-2}$ for
$m = 10^{-2}$\,eV due to the linear $m$ dependence.

At small redshift, the curves have asymptotic values
\begin{subequations}
\begin{eqnarray}
  \langle \rho_-(z\rightarrow 0) \rangle
& \approx &
  \frac {\hat \rho^2_0}
        {(1 + \hat \rho^2_0)}
\approx
  2.8 \times (10^{-4}, 10^{-2}),
\\
  \langle \rho_R(z\rightarrow 0) \rangle
& \approx &
  \frac {\hat \rho_0}
        {(1 + \hat \rho^2_0)}
\approx
  1.7 \times (10^{-2}, 10^{-1}),
\qquad
\end{eqnarray}
\end{subequations}
for $m = (10^{-2}, 10^{-3})$\,eV, respectively.
This explains the flat behavior of the curves in
the small $z$ region.
For $m = |{\bf p}_0| \sim 1.66\times 10^{-4}$ eV,
the asymptotic values are $\langle \rho_R(z\rightarrow
0)\rangle = 1-\langle\rho_-(z\rightarrow 0)\rangle = 1/2$.
Any value between $[0,1/2]$ can be obtained if
$m< |{\bf p}_0|$ and the vanishing value $0$ occurs
for $m\rightarrow 0$, where no transition between
the left- and right-chiral components can occur.

{\it Detection of Cosmic Neutrinos}
--
The chiral oscillation has important consequences
on the chiral composition of relic neutrinos.
To make a uniform description for the Dirac and Majorana
cases, we denote the four
chiral and helicity components of neutrino/antineutrino
as $\nu^\pm_c$, with $c = L,R$ representing the chirality
and $\pm$ the helicity, respectively. Although all
these four states can in
principle be created even with just left-chiral current
interactions, the right-chiral neutrino ($\nu^+_L$) and left-chiral
antineutrino ($\nu^-_R$) are highly suppressed
by a factor of 
$m^2 / E^2 < 10^{-12}$ due to
the high temperature and hence large neutrino momentum
at the decoupling time
and therefore can be safely ignored \cite{Long:2014zva}.
Nevertheless, these missing states ($\nu^+_L$ and
$\nu^-_R$) will become populated
through chiral oscillation as elaborated above.
This can change the event number of cosmic neutrino
background detection at the PTOLEMY
experiment \cite{Betti:2019ouf}.

At production, neutrinos are in thermal equilibrium
and hence follow the Fermi-Dirac distribution with
number densities $n(\nu^-_L) = n(\nu^+_R) = n(z_d)$
while $n(\nu^+_L) = n(\nu^-_R) = 0$.
In the presence of chiral oscillation, the original
number densities $n(\nu^-_L)$ and $n(\nu^+_R)$ of
neutrino and antineutrino, respectively,
splits into left/right-chiral components. To be more
specific, the averaged solution \geqn{eq:rho_1nu}
indicates that the neutrino number density is
redistributed as,
\begin{equation}
  n(\nu^-_L)
=
  n_0
\left(
  1
- \frac{m^2}{2E^2}
\right),
\quad {\rm and} \quad
  n(\nu^-_R)
=
  n_0 \frac{m^2}{2E^2}.
\end{equation}
Similar things happen for the anti-neutrino mode.
The full results are summarized in the left
half of \gtab{tab:numb_density}. Since the same chiral
oscillation properties apply for both Dirac and
Majorana neutrinos, the neutrino number densities in
\gtab{tab:numb_density} also apply universally.

For comparison, we also show the case without
considering chiral oscillation. After decoupling,
the number densities are then fixed and decreases
with $a^3$ if we do not take into account neutrino
chiral oscillation. From the neutrino decoupling
($z_d \sim 6 \times 10^9$) to the present time ($z=0$),
the temperature drops from 1\,MeV to $0.168$\,meV,
resulting in $n_0 \equiv n(z = 0) = 56$\,cm$^{-3}$
\cite{Long:2014zva}. Without considering chiral
oscillation, there is no need to distinguish different
chiralities, but only $n(\nu^\pm)$. The results have
been summarized in the right half of
\gtab{tab:numb_density}.

Since relic neutrinos are nonrelativistic,
the most feasible way of detecting them is through
neutrino capture on tritium ($T$) \cite{Weinberg:1962zza,Betti:2019ouf}.
The squared matrix element is,
\begin{align}
 |\mathcal{M}|^2
=
  \langle ^3{\rm He}\,e^- | \mathcal L_{\rm int} |T\nu\rangle 
  \langle \nu T| \mathcal L_{\rm int} | ^3{\rm He}\,e^- \rangle.
\label{eq:nu_matrix2}
\end{align}
The quantity $|\nu\rangle \langle \nu|$ encodes the
density matrix in \geqn{eq:density_matrix}.
Another way of seeing this quantity is through the
evolved spinor in \geqn{eq:spinor_LH_time}.
In the usual charged-current interaction,
$\mathcal L_{\rm int} \equiv \frac{g}{\sqrt{2}} W^-_\mu 
  \overline{\ell} \gamma^\mu  P_L \nu$,
the second part gives an extra detection possibility
only for Majorana neutrinos.

Due to parity violation, only the left-chiral
neutrino and the right-chiral anti-neutrino
appear in the interaction. In addition,
energy conservation selects neutron in the tritium
nuclei as the initial and proton as the final states.
Then, charge conservation
selects electron in the final state. For relativistic
neutrinos, only the left-chiral neutrino can be detected.
At low energy, only neutrino can contribute for the Dirac
case while the two helicity states have equal contributions
for the Majorana case,
\begin{subequations}
\begin{eqnarray}
  N_{\nu_D}
& \hspace{-1mm} = &
  \bar \sigma N_T
\hspace{-1mm}
\left( 1 + \frac{|{\bf p}|} E \right)
  n(\nu^-_L),
\\ 
    N_{\nu_M}
& \hspace{-1mm} = &
  \bar \sigma N_T
\hspace{-0.5mm}
\left[
  \left( 1 + \frac{|{\bf p}|} E \right) n(\nu_L^-)
\hspace{-0.5mm} + \hspace{-0.5mm}
  \left( 1 - \frac{|{\bf p}|} E \right) n(\nu_R^+)
\right]\hspace{-0.5mm},
\qquad
\end{eqnarray}
\end{subequations}
where $N_T$ is the number of targets in the detector.
The factors $1 \pm |{\bf p}|/E$ are associated with helicities,
$h = \mp$, respectively, and the cross section
$\bar \sigma$ can be found in \cite{Long:2014zva}.
\begin{table}[t]
\centering
\begin{tabular}{c|cc||c|cc}
& \multicolumn{2}{c||}{w/ Chiral Osc.} & & \multicolumn{2}{c}{w/o Chiral Osc.}
\\
\hline\hline
& $z = z_{\rm fo}$ & $z = 0$ & & $z = z_{\rm fo}$ & $ z = 0$
\\
\hline
$n(\nu_L^-)$ & $n_0(1+z_{\rm fo})^3$ & $n_0\left(1-\frac{m^2}{2E^2}\right)$ &\multirow{2}{*}{$n(\nu^-)$} & \multirow{2}{*}{$n_0(1+z_{\rm fo})^3$} & \multirow{2}{*}{$n_0$}

\\
$n(\nu_R^-)$ & 0 & $n_0 \frac {m^2}{2 E^2}$ & & & 
\\
$n(\nu_L^+)$ & 0 & $n_0 \frac {m^2}{2 E^2}$ & \multirow{2}{*}{$n(\nu^+)$} & \multirow{2}{*}{$n_0(1+z_{\rm fo})^3$} & \multirow{2}{*}{$n_0$}
\\
$n(\nu_R^+)$ & $n_0(1+z_{\rm fo})^3$ & $n_0\left(1-\frac{m^2}{2E^2}\right)$ & & &
\\
\end{tabular}
\caption{
The neutrino number densities at neutrino decoupling
($z_d \approx 6\times 10^9$) and today ($z= 0$) for all
the four different chirality and helicity components
($\nu^-_L$, $\nu^-_R$, $\nu^+_R$, and $\nu^-_R$).
The number density $n_0 = 56$ cm$^{-3}$ is obtained
according to the Fermi-Dirac distribution for the
neutrino temperature $T_\nu \approx 0.168$\,meV
\cite{Long:2014zva}.}
\label{tab:numb_density}
\end{table}
Without chiral oscillation, $n(\nu^-) = n(\nu^+) = n_0$,
leading to the ratio between the Majorana/Dirac event rates as:
$N_{\nu_M} / N_{\nu_D} \approx 2$. 
In the presence of chiral oscillation,
the chiral components experience frequent
oscillation to and from each other.
The averaged number densities have been summarized in
\gtab{tab:numb_density}.
In the nonrelativistic limit, $|{\bf p}|\rightarrow 0$,
$n(\nu_L^-) = n(\nu_R^+) = n(\nu_L^+) = n(\nu_R^-) = n_0/2$.
The expected {\cnb} event rates are reduced to half of the
conventional prediction
without chiral oscillation.

{\it Conclusions}
--
We explore the phenomenological consequences of parity
violation and chirality mixture of the neutrino mass
term on neutrino oscillation for both Dirac and Majorana
neutrino. Even with only one flavor, the left- and
right-chiral components can already oscillate
to and from each other. This has a significant
consequence on the detection of the cosmic neutrino
background. Although helicity is conserved in chiral
oscillation, the neutrino chirality changes and the
neutrino number density equality splits into the two
chiral components. This reduces the event rate of the
{\cnb} detection by a factor of 2. Once knowing whether
neutrinos are Dirac or Majorana fermions, we can test
chiral oscillation with {\cnb} detection.

{\it Notes Added}
--
During the finalization of this paper, we notice
a similar work \cite{Bittencourt:2020xen} about
the phenomenological consequences of chiral oscillation
on the cosmological relic neutrinos.

{\it Acknowledgement}
--
SFG is grateful to the Double First Class start-up
fund (WF220442604) provided by Tsung-Dao Lee Institute and
Shanghai Jiao Tong University. SFG would like to thank
Yu-Feng Li for useful discussions.

\end{document}